\documentclass[twocolumn,pra,showpacs,superscriptaddress,amssymb,amsmath,amsmath]{revtex4-1}
\usepackage{graphicx}
\usepackage{epstopdf}
\usepackage{bm}
\usepackage{hyperref}
\usepackage{comment}

\hypersetup{%
   pdfpagemode=None, %FullScreen,
   pdfstartpage=1,
   pdfmenubar=true,
   pdftoolbar=true,
   colorlinks = true,
   linkcolor=blue,
   citecolor=blue,
   urlcolor=blue,
   bookmarksopen=false
 }

\newcommand{\be}{\begin{equation}}
\newcommand{\ee}{\end{equation}}

\begin{document}
\title{Cold interactions and chemical reactions of linear polyatomic anions \\ with alkali-metal and alkaline-earth-metal atoms}

\author{Micha\l~Tomza}
\email{m.tomza@cent.uw.edu.pl}
\affiliation{Centre of New Technologies, University of Warsaw, Banacha 2c, 02-097 Warsaw, Poland}

\date{\today}

\begin{abstract}
We consider collisional studies of linear polyatomic ions immersed in ultracold atomic gases and investigate the intermolecular interactions and chemical reactions of several molecular anions (OH$^-$, CN$^-$, NCO$^-$, C$_2$H$^-$, C$_4$H$^-$) with alkali-metal (Li, Na, K, Rb, Cs) and alkaline-earth-metal (Mg, Ca, Sr, Ba) atoms. State-of-the-art \textit{ab initio} techniques are applied to compute the potential energy surfaces (PESs) for these systems.
The coupled cluster method restricted to single, double, and noniterative triple excitations, CCSD(T), is employed and the scalar
relativistic effects in heavier metal atoms are modeled within the small-core energy-consistent pseudopotentials.
The leading long-range isotropic and anisotropic induction and dispersion interaction coefficients are obtained within the perturbation theory.
The PESs are characterized in detail and their universal similarities typical for systems dominated by the induction interaction are discussed. 
The two-dimensional PESs are provided for selected systems and can be employed in scattering calculations.
The possible channels of chemical reactions and their control are analyzed based on the energetics of reactants. The present study of the electronic structure is the first step towards the evaluation of prospects for sympathetic cooling and controlled chemistry of linear polyatomic ions with ultracold atoms.
\end{abstract}


\maketitle

\section{Introduction}

Experiments at low and ultralow temperatures allow investigating physics and chemistry at the fundamental quantum limit~\cite{ColdAtomsandMolecules}. The realization of ultracold atomic gases have significantly increased our understanding of quantum many-body systems~\cite{BlochRMP08}. The production of ultracold gases of polar diatomic molecules have resulted in groundbreaking experiments on controlled chemical reactions in the quantum regime~\cite{QuemenerCR12}. The hybrid systems of laser-cooled trapped ions combined with ultracold atoms in a single experimental setup have also recently become a new platform for investigating quantum matter~\cite{HarterCP14}. Polyatomic molecules and molecular ions have additional rotational and vibrational degrees of freedom that could potentially be used for various applications~\cite{DulieuPCCP11}, therefore first experiments on cooling of trapped polyatomic molecules have been launched~\cite{BethlemNature00,ZeppenfeldNature12,GlocknerPRL15,PrehnPRL16,KozyryevPRL17}.

The hybrid systems of laser-cooled trapped ions immersed in ultracold atomic gases~\cite{HarterCP14} combine the best features of the two well-established fields of research: ultracold atoms~\cite{BlochRMP08} and trapped ions~\cite{LeibfriedRMP03}. Their potential applications range from cold controlled ion-atom collisions and chemical reactions~\cite{ZipkesPRL10,ZipkesNature10,HallPRL11,RatschbacherNatPhys12} to quantum simulations of solid-state physics~\cite{GerritsmaPRL12,BissbortPRL13} and quantum computations~\cite{DoerkPRA10}. Cold molecular ions can be formed from cold mixtures of atomic ions and atoms~\cite{SullivanPCCP11,TomzaPRA15a,TomzaPRA15b} or they can be cooled down from room temperature using laser, buffer-gas, or sympathetic cooling~\cite{RellergertNature13,HansenNature14,StoecklinNatCom16}. Potential applications of molecular ions include precision measurements~\cite{GermannNatPhys14,LohScience13}, cold controlled chemistry~\cite{WillitschPCCP08,MikoschIRPC10}, and novel quantum simulations~\cite{MidyaPRA16}. 

Recently, first experiments combining simple diatomic molecular ions with ultracold atoms have been launched. 
The dynamics of N$_2^+$ molecular cations immersed in ultracold Rb atoms~\cite{HallPRL12} and BaCl$^+$ cations immersed in ultracold Ca atoms~\cite{RellergertNature13,StoecklinNatCom16} was investigated.
The collisions of OH$^-$ molecular anions with ultracold Rb atoms were studied both experimentally~\cite{DeiglmayrPRA12,GonzalezCP15} and theoretically~\cite{GonzalezEPJD08,TacconiJCP09,ByrdPRA13,GonzalezCP15,GonzalezNJP15}.
The cooling of simple molecular ions such as MgH$^+$, NH$_2^-$, and OH$^-$ immersed in cold buffer gases of helium or molecular  hydrogen  were also experimentally~\cite{HansenNature14,OttoPRL08,GerlichFD09,HauserNatPhys15,MulinPCCP15,HauserNJP15} and theoretically~\cite{BodoPRL02,StoecklinJCP11,EndresJPCA14} investigated. 
Unfortunately, there is very little knowledge of cold and ultracold interactions, collisions, and reactions between polyatomic molecular ions and alkali-metal or alkaline-earth-metal atoms at the moment, hence prospects for sympathetic cooling of polyatomic molecular ions down to low and ultralow temperatures are not known. % limiting advances in experimental investigations of these systems. 

In the present work, we investigate the intermolecular interactions of several molecular anions (OH$^-$, CN$^-$, NCO$^-$, C$_2$H$^-$, C$_4$H$^-$) with alkali-metal (Li, Na, K, Rb, Cs) and alkaline-earth-metal (Mg, Ca, Sr, Ba) atoms.
We calculate and characterize the potential energy surfaces, long-range induction and dispersion interaction coefficients, and possible channels of chemical reactions and their control by using state-of-the-art \textit{ab initio} techniques.
Selected diatomic molecular anions are important in many areas of chemistry, whereas considered polyatomic molecular anions are of great interest to astrochemistry~\cite{LarssonRPR12}.
Surprisingly, a number of cations but just a few anions have been conclusively detected in the interstellar space. At the moment, six anions confirmed in the interstellar medium are CN$^-$, C$_4$H$^-$, C$_6$H$^-$, C$_8$H$^-$, C$_3$N$^-$, C$_5$N$^-$~\cite{HerbstPCCP14,FortenberryJPCA15}. The NCO$^-$ anion was also detected but is believed to be trapped in astronomical ices~\cite{HudsonAJ01}. 
The conditions in ultracold ion-atom experiments with alkali-metal and alkaline-earth-metal atoms are more extreme and unlike those that occur in the interstellar space. Nevertheless, precision spectroscopy of cold polyatomic molecular anions and investigations of their stability, properties of valence and dipole-bound excited states, as well as detailed studies of cold chemical reactions and their mechanisms can { potentially} shed new light on the chemistry of anions in the universe~\cite{HerbstPCCP14}.

The selected molecular anions have relativity simple closed-shell electronic structure, large dipole moments, and high binding energies~\cite{DouguetJCP15}, which makes them convenient candidates both for theoretical and experimental studies. 
These and similar molecular ions were also already spectroscopically investigated in ion traps~\cite{WesterJPB09}, therefore the experimental realization of considered hybrid systems should be feasible. 
The results for diatomic molecular anions can serve as a benchmark and reference for studies of polyatomic anions. 

The plan of this paper is as follows. Section~\ref{sec:theory} describes the theoretical methods used in the \textit{ab initio} electronic structure calculations. Section~\ref{sec:results} presents and discusses the properties of considered molecular ions and atoms, the potential energy surfaces, and the leading long-range induction and dispersion interaction coefficients. The prospects for chemical reactions and their control are also analyzed. Section~\ref{sec:summary} summarizes our paper and presents future possible applications.

\section{Computational details}
\label{sec:theory}

The electronic ground state of the OH$^-$, CN$^-$, NCO$^-$, C$_2$H$^-$, and C$_4$H$^-$ molecular anions is of the singlet $^1\Sigma^+$ symmetry, thus these ions are closed-shell and can be accurately described with \textit{ab initio} electronic structure methods of quantum chemistry. The interaction between a closed-shell $^1\Sigma^+$-state molecular anion and an open-shell $^2S$-state alkali-metal atom (a closed-shell $^1S$-state alkaline-earth-metal atom) results in one electronic state of the ${}^2A'$ (${}^1A'$) symmetry. 
In this paper, working within the Born-Oppenheimer approximation, we describe linear molecular anions within the rigid rotor approximation and use Jacobi coordinates to describe the relative orientation of a molecular anion and an atom.
Therefore, the potential energy surfaces (PESs) are functions of two coordinates $V(R,\theta)$, where $R$ is the distance between an atom and the center of mass of a molecule anion, and $\theta$ is the angle between the axis of a molecular anion (oriented from a heavier atom to a lighter one) and the axis connecting an atom with the center of mass of a molecular anion (oriented from a molecular anion to an atom).

In order to obtain potential energy surfaces, we adopt the computational scheme successfully applied to the ground-state interactions between polar alkali-metal dimer~\cite{TomzaPRA13b}, an ytterbium cation with a lithium atom~\cite{TomzaPRA15a}, a chromium atom with alkaline-earth-metal atoms~\cite{TomzaPRA13a}, and an europium atom with alkali-metal and alkaline-earth-metal atoms~\cite{TomzaPRA14}.
Thus, to calculate PESs for anions interacting with alkaline-earth-metal atoms (alkali-metal atoms) we employ the close-shell (spin-restricted open-shell) coupled cluster method restricted to single, double, and noniterative triple excitations, starting from the restricted close-shell (open-shell) Hartree-Fock orbitals, CCSD(T)~\cite{PurvisJCP82,KnowlesJCP93}.
The interaction energies are obtained with the supermolecule method and
the basis set superposition error is corrected by using the counterpoise correction~\cite{BoysMP70}
\begin{equation}
V_{\textrm{ion+atom}}=E_\textrm{ion+atom}-E_{\textrm{ion}}-E_\textrm{atom}\,,
\end{equation}
where $E_\textrm{ion+atom}$ denotes the total energy of a molecular ion interacting with an atom, and $E_\textrm{ion}$ and $E_\textrm{atom}$ are the total energies of a molecular ion and an atom computed in a dimer basis set. 

The Li, Na, and Mg atoms are described with the augmented correlation-consistent polarized core-valence quadruple-$\zeta$ quality basis sets (aug-cc-pCVQZ)~\cite{PrascherTCA10}, whereas the H, C, N, and O atoms are described with the aug-cc-pVQZ~\cite{DunningJCP89,KendallJCP92} basis sets in calculations of intermolecular interaction and with the aug-cc-pCV5Z basis sets~\cite{DunningJCP89,KendallJCP92} in calculations of molecular properties.
The scalar relativistic effects in the K, Rb, Cs, Ca, Sr and Ba atoms are included by employing the small-core relativistic energy-consistent pseudopotentials (ECP) to replace the inner-shells electrons~\cite{DolgCR12}. The use of the pseudopotentials allows one to use larger basis sets to describe the valence electrons and models the inner-shells electrons density as accurately as the high quality atomic calculation used to fit the pseudopotentials.
The pseudopotentials from the Stuttgart library are employed in all calculations. 
The K, Ca, Rb, Sr, Cs, and Ba atoms are  described with the ECP10MDF, ECP10MDF, ECP28MDF, ECP28MDF, ECP46MDF, and ECP46MDF pseudopotentials~\cite{LimJCP06,DolgTCA98} and the $[11s11p5d3f]$, $[12s12p7d4f2g]$, $[14s14p7d6f1g]$, $[14s11p6d5f4g]$, $[12s11p6d4f2g]$, and $[13s12p6d5f4g]$ basis sets, respectively, obtained by decontracting and augmenting the basis sets suggested in Refs.~\cite{LimJCP06,DolgTCA98}. 
The used basis sets were optimized in Refs.~\cite{TomzaPCCP11,TomzaMP13,TomzaPRA13a,TomzaPRA14}.
The basis sets are additionally augmented in all calculations by the set of the $[3s3p2d1f1g]$ bond functions~\cite{midbond}.

The potential energy surfaces from the molecular body-fixed calculations $V(R,\theta)$ can be expanded into the basis of the Legendre polynomials $P_\lambda(\cdot)$~\cite{BuckinghamCR88}
\begin{equation}\label{eq:Vn}
V(R,\theta)=\sum_{\lambda=0}^{\lambda_{max}-1} V_{\lambda}(R) P_{\lambda}(\cos\theta) \,.
\end{equation}
Such a decomposition is especially convenient for coupled channels scattering calculations~\cite{AlthorpeARPC03}.
Here, we calculate the potential energy surfaces $V(R,\theta)$ on the two-dimensional grid consisting of around 25 points in the ion-atom distance $R$ with values between around 2.5$\,$bohr and 30$\,$bohr and 12 points in the angle $\theta$ with values between 0 and 180 degrees chosen to be the quadratures for the Legendre polynomial of the order $\lambda_{max}=12$.
The Legendre components $V_\lambda(R)$ are obtained by integrating out \textit{ab initio} points.

The intermolecular interaction energy between a linear closed-shell polar molecular ion and a $S$-state atom, both in the electronic ground state, at large intermolecular distances $R$, in the molecular frame, is of the form~\cite{HeijmenMP96}
\begin{equation}\label{eq:long-range}
\begin{split}
V(R,\theta)\approx&-\frac{C^\textrm{ind}_{4}}{R^4}-\frac{C^\textrm{ind}_{5,1}}{R^5}\cos\theta-\frac{C^\textrm{ind}_{6,0}}{R^6}-\frac{C^\textrm{disp}_{6,0}}{R^6}\\&-\left(\frac{C^\textrm{ind}_{6,2}}{R^6}+\frac{C^\textrm{disp}_{6,2}}{R^6}\right)P_2(\cos\theta)+\dots\,,
\end{split}
\end{equation}  
and the Legendre components of Eq.~\eqref{eq:Vn} are
\begin{equation}\label{eq:long}  
\begin{split}
V_0(R)\approx&-\frac{C^\textrm{ind}_{4}}{R^4}-\frac{C^\textrm{ind}_{6,0}}{R^6}-\frac{C^\textrm{disp}_{6,0}}{R^6}+\dots\\\,
V_1(R)\approx&-\frac{C^\textrm{ind}_{5,1}}{R^5}+\dots\,\\
V_2(R)\approx&-\frac{C^\textrm{ind}_{6,2}}{R^6}-\frac{ C^\textrm{disp}_{6,2}}{R^6}+\dots\,.
\end{split}
\end{equation} 
Proper treatment of the interaction potential at large internuclear distances is especially important for ultracold collisions. Different $V_\lambda$ terms govern inelastic rotational transitions, changing molecular rotation by $\Delta j =\pm\lambda$.   

The leading long-range induction coefficients are 
\begin{equation}\label{eq:Cnind}
\begin{split}
C^{\mathrm{ind}}_{4}&=\frac{1}{2}q^2\alpha_\textrm{atom}\,, \\
C^{\mathrm{ind}}_{5,1}&=2d_\textrm{ion}q\alpha_\textrm{atom}\,,  \\
C^{\mathrm{ind}}_{6,0}&=\frac{1}{2}q^2\beta_\textrm{atom}+d^2_\textrm{ion}\alpha_\textrm{atom} \,,\\
C^{\mathrm{ind}}_{6,2}&=2\Theta_\textrm{ion}q\alpha_\textrm{atom}+d^2_\textrm{ion}\alpha_\textrm{atom}\,,\\
\end{split}
\end{equation}  
where $q$ is the charge of the molecular ion, $\alpha_\textrm{atom}$ is the static electric dipole polarizability of the atom, $d_\textrm{ion}$ is the permanent electric dipole moment of the molecular ion, $\Theta_\textrm{ion}$ is the permanent electric quadruple moment of the molecular ion, and $\beta_\textrm{atom}$ is the static electric quadrupole polarizability of the atom. 
The leading long-range dispersion coefficients are
\begin{equation}\label{eq:Cndisp}
\begin{split}
C^\textrm{disp}_{6,0}&=\frac{3}{\pi}\int_0^\infty  \bar{\alpha}_{\textrm{ion}}(i\omega){\alpha}_\textrm{atom}(i\omega)d\omega\,,\\
C^\textrm{disp}_{6,2}&=\frac{1}{\pi}\int_0^\infty \Delta\alpha_{\textrm{ion}}(i\omega){\alpha}_\textrm{atom}(i\omega)d\omega \,,
\end{split}
\end{equation}  
where ${\alpha}_{\textrm{atom(ion)}}(i\omega)$ is the dynamic polarizability of the atom(ion) at imaginary frequency and the
average polarizability and polarizability anisotropy are given by $\bar{\alpha}=(\alpha_\parallel+2\alpha_\perp)/3$ and $\Delta\alpha=\alpha_\parallel-\alpha_\perp$, respectively, with $\alpha_\parallel$ and $\alpha_\perp$ being the components of the polarizability tensor parallel and perpendicular to the internuclear axis of the molecular ion.

The equations~\eqref{eq:long-range}-\eqref{eq:Cndisp} result from the long-range multipole expansion of the intermolecular interaction energy within the perturbation theory, therefore different terms are given by electric properties of monomers~\cite{HeijmenMP96}.  
The $-C_4^\mathrm{ind}/R^4$ term describes the interaction between the charge of the molecular ion and the induced electric dipole moment of the atom. The $-C_{5,1}^\mathrm{ind}/R^5\cos\theta$ term describes the interaction between the permanent electric dipole moment of the molecular ion and the induced electric dipole moment of the atom. The first term in $-C_{6,0}^\mathrm{ind}/R^6$ describes the interaction between the charge of the molecular ion and the induced electric quadruple moment of the atom, whereas the second one describes the interaction between the permanent electric dipole moment of the molecular ion and the higher-order induced electric dipole moment of the atom. The first term in $-C_{6,2}^\mathrm{ind}/R^6$ describes the interaction between the permanent electric quadruple moment of the molecular ion and the induced electric dipole moment of the atom, and the second one is the same as in $-C_{6,0}^\mathrm{ind}/R^6$. The dispersion terms result from the interaction between instantaneous dipole-induced dipole moments of the molecular ion and atom arising due to quantum fluctuations.   

The dynamic electric dipole polarizabilities at imaginary frequency $\alpha(i\omega)$ of alkali-metal and alkaline-earth-metal atoms are taken from Ref.~\cite{DerevienkoADNDT10}, whereas the dynamic polarizabilities of molecular anions are obtained by using the explicitly connected representation of the expectation value and polarization propagator within the coupled cluster method~\cite{MoszynskiCCCC05} and the best approximation proposed in Ref.~\cite{KoronaMP06}. 

The static electric dipole and quadrupole polarizabilities of atoms and the permanent electric dipole and quadrupole moments of molecular anions are calculated with the CCSD(T) and finite field methods. 

All electronic structure calculations are performed with the \textsc{Molpro} package of \textit{ab initio} programs \cite{Molpro}.

\section{Numerical results and discussion}
\label{sec:results}

\begin{figure}[tb]
\begin{center}
\includegraphics[width=\columnwidth]{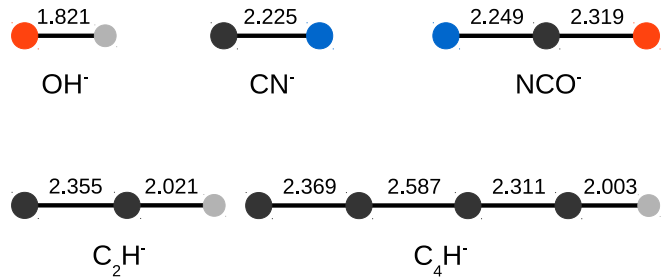}
\end{center}
\caption{The equilibrium geometries of  molecular anions calculated with the CCSD(T) method and aug-cc-pCV5Z basis sets. Bond distances are in bohr. The numerical uncertainty is $\pm$0.002$\,$bohr.}
\label{fig:geom}
\end{figure}

\subsection{Properties of molecular anions and atoms}

An accurate description of monomers and reproduction of their properties are essential for a proper evaluation of intermolecular interactions and chemical reactions.
Therefore, in this subsection, we examine the electronic properties of investigated molecular anions and atoms, which also define the long-range interaction coefficients crucial for cold physics and chemistry.

The \textit{ab initio} description of anions is usually more challenging as compared to calculations involving cations. Specifically, such calculations require basis sets with diffuse functions to account for the expanded character of anionic valence and dipole-bound ground and excited electronic states~\cite{RienstraCR02}. Therefore, we use the augmented polarized basis sets of at least quadruple-$\zeta$ quality which combined with relatively large binding energies of selected closed-shell molecular anions should provide accurate results.

\begin{table}[t!]
\caption{Characteristics of molecular anions at the equilibrium geometry: permanent electric dipole moment $d_e$, permanent electric quadrupole moment $\Theta_e$, parallel $\alpha_e^\parallel$ and perpendicular  $\alpha_e^\perp$ components of the static electric dipole polarizability,  and vertical  electron detachment energy ED. The results for two basis sets are presented: A - aug-cc-pVQZ and B - aug-cc-pCV5Z. \label{tab:anions}}  
\begin{ruledtabular}
\begin{tabular}{llrrrrr}
Ion & Set  & $d_e\,$(D) &  $\Theta_e\,$(a.u.) & $\alpha_e^\parallel\,$(a.u.) & $\alpha_e^\perp\,$(a.u.) & ED$\,$(cm$^{-1}$)\\
\hline
OH$^-$     & A  &   1.08 &  3.03 & 22.0 & 30.1 & 14505 \\
           & B  &   1.07 &  3.09 & 23.2 & 32.6 & 14678 \\
CN$^-$     & A  & -0.652 & -5.50 & 37.3 & 27.6 & 31329 \\
           & B  & -0.655 & -5.48 & 37.2 & 27.7 & 31543 \\               
NCO$^-$    & A  &  -1.54 & -12.8 & 44.8 & 24.7 & 30452 \\
           & B  &  -1.52 & -12.8 & 45.0 & 24.8 & 30718 \\
C$_2$H$^-$ & A  &   3.22 & -2.20 & 55.8 & 40.8 & 24478 \\
           & B  &   3.22 & -2.18 & 55.5 & 41.0 & 24587 \\
C$_4$H$^-$ & A  &   6.19 & -18.5 & 117.0 & 46.7 & 29892 \\
           & B  &   6.20 & -18.4 & 116.6 & 46.9 & 30026 \\
\end{tabular}
\end{ruledtabular}
\end{table}
\begin{table}[t!]
\caption{Characteristics of alkali-metal and alkaline-earth-metal atoms: static electric dipole polarizability $\alpha$, static electric quadrupole polarizability $\beta$, ionization potential IP, electron affinity EA, and the lowest $S$--$P$ excitation energy (${}^2S$--${}^2P$ for alkali-metal atoms and ${}^1S$--${}^3P$ for alkaline-earth-metal atoms). Present theoretical values are compared with the most accurate available experimental or theoretical data.\label{tab:atoms}}  
\begin{ruledtabular}
\begin{tabular}{llllll}
Atom & $\alpha\,$(a.u.) & $\beta\,$(a.u.) & IP$\,$(cm$^{-1}$) & EA$\,$(cm$^{-1}$) & $S$--$P$$\,$(cm$^{-1}$)\\
\hline
Li &  164.3 &  1414 & 43464 &  4970 & 14911 \\
   &  164.2\cite{MiffreEPJ06} & 1423\cite{YanPRA96} & 43487\cite{nist} &  4885\cite{AndersenJPCRD99} & 14904\cite{nist} \\
Na &  166.4 &  1920 & 41217 &  4406 & 16799 \\
   &  162.7\cite{EkstromPRA95} & 1895\cite{KaurPRA15} & 41449\cite{nist} &  4419\cite{AndersenJPCRD99} &  16968\cite{nist}     \\
K  &  290.8 &  4970 & 34949 &  4015 & 13022 \\
   &  290.0\cite{GregoirePRA15}  & 4947\cite{KaurPRA15} & 35010\cite{nist} &  4045\cite{AndersenJPCRD99} & 13024\cite{nist}      \\
Rb &  319.5 &  6578 & 33566 &  3887 & 12686 \\
   &  320.1\cite{GregoirePRA15} &  6491\cite{KaurPRA15} & 33691\cite{nist} &  3919\cite{AndersenJPCRD99} &  12737\cite{nist}     \\
Cs &  395.5 & 10343 & 31331 &  3728 & 11594 \\
   &  401.2\cite{GregoirePRA15}   & 10470\cite{PorsevJCP03} & 31406~\cite{nist} &  3804\cite{AndersenJPCRD99} &  11548\cite{nist}     \\
Mg &   71.8 &   821 & 61466 &  -1824 & 21701 \\
   &   71.3\cite{PorsevJETP}  &  812\cite{PorsevJETP} & 61671\cite{nist} &  $<0$\cite{AndersenJPCRD99} & 21891\cite{nist}     \\
Ca &  156.9 &  2946 & 49243 &  -491 & 15190 \\
   &  157.1\cite{PorsevJETP}  &  3081\cite{PorsevJETP} & 49306\cite{nist} &  198\cite{AndersenJPCRD99} & 15263\cite{nist}      \\
Sr &  199.2 &  4551 & 45814 &  -4.3 & 14639 \\
   &  197.2\cite{PorsevJETP} &  4630\cite{PorsevJETP} & 45932\cite{nist} &   420\cite{AndersenJPCRD99} & 14705\cite{nist}   \\
Ba &  276.8 &  8586 & 41780 &   580 & 13106 \\
   &  273.5\cite{PorsevJETP} &  8900\cite{PorsevJETP} & 42035\cite{nist} &  1166\cite{AndersenJPCRD99} &  13099\cite{nist}  \\
\end{tabular}
\end{ruledtabular}
\end{table}

\begin{table*}[tbh!]
\caption{Characteristics of the potential energy surfaces for molecular anions interacting with alkali-metal and alkaline-earth-metal atoms, all in the electronic ground state: equilibrium intermolecular distance $R_e$ and well depth $D_e$ for the two linear geometries ($C_{2v}$ symmetry) corresponding to global minimum, local minimum, or saddle point, and induction $C_{n,k}^{\mathrm{ind}}$ and dispersion $C_{n,k}^{\mathrm{disp}}$ coefficients describing the long-range part of the interaction. Long-range coefficients are in atomic units.\label{tab:int}} 
\begin{ruledtabular}
\begin{tabular}{lrrrrrrrrrr}
System & $R_e\,$(bohr) & $D_e\,$(cm$^{-1}$) & $R_e'\,$(bohr) & $D_e'\,$(cm$^{-1}$) & $C^{\mathrm{ind}}_{4}\,$ & $C^{\mathrm{ind}}_{5,1}\,$ & $C^{\mathrm{ind}}_{6,0}\,$ & $C^{\mathrm{ind}}_{6,2}\,$ & $C^{\mathrm{disp}}_{6,0}\,$ & $C^{\mathrm{disp}}_{6,2}\,$\\
\hline
OH$^-$+Li & 3.18 & 23917 & 4.85 &  6672 &  82.1 &  -139  &  736 &  -987 &  387 & -22.8 \\
OH$^-$+Na & 3.91 & 16700 & 5.44 &  5246 &  83.2 &  -141  &  990 & -1000 &  426 & -23.5 \\
OH$^-$+K  & 4.44 & 16693 & 5.86 &  6314 & 145.4 &  -246  & 2308 & -1747 &  648 & -38.1 \\
OH$^-$+Rb & 4.62 & 16498 & 6.04 &  6377 & 159.8 &  -270  & 3346 & -1919 &  721 & -41.3 \\
OH$^-$+Cs & 4.82 & 17138 & 6.24 &  6904 & 197.7 &  -334  & 5242 & -2376 &  868 & -49.6 \\
OH$^-$+Mg & 3.58 & 20283 & 5.54 &  3792 &  35.9 & -60.6  &  423 &  -431 &  312 & -11.1 \\
OH$^-$+Ca & 3.98 & 25514 & 5.63 &  6833 &  78.5 &  -133  & 1501 &  -942 &  548 & -24.0 \\
OH$^-$+Sr & 4.21 & 25458 & 5.79 &  7581 &  99.6 &  -168  & 2311 & -1197 &  658 & -29.7 \\ 
OH$^-$+Ba & 4.40 & 27552 & 5.91 &  9246 & 138.4 &  -234  & 4342 & -1663 &  832 & -39.6 \\
\hline
CN$^-$+Li & 4.54 & 14994 & 5.01 & 14224 &  82.1 & 84.6 &  718 & 1813 &  436 & 43.7 \\
CN$^-$+Na & 5.26 & 10325 & 5.71 &  9960 &  83.2 & 85.7 &  971 & 1836 &  484 & 48.2 \\
CN$^-$+K  & 5.90 & 10045 & 6.40 &  9540 & 145.4 &  150 & 2275 & 3208 &  732 & 72.7 \\
CN$^-$+Rb & 6.14 &  9625 & 6.66 &  9096 & 159.8 &  165 & 3310 & 3525 &  817 & 80.9 \\
CN$^-$+Cs & 6.42 &  9732 & 6.95 &  9150 & 197.7 &  204 & 5198 & 4363 &  985 & 97.2 \\
CN$^-$+Mg & 4.98 &  9793 & 5.49 &  8653 &  35.9 & 37.0 &  415 &  792 &  372 & 36.1 \\
CN$^-$+Ca & 5.41 & 13920 & 5.93 & 12571 &  78.5 & 80.8 & 1483 & 1731 &  641 & 62.8 \\
CN$^-$+Sr & 5.68 & 13942 & 6.20 & 12650 &  99.6 &  103 & 2289 & 2198 &  767 & 75.1 \\
CN$^-$+Ba & 5.91 & 15281 & 6.43 & 14038 & 138.4 &  143 & 4311 & 3054 &  965 & 94.6 \\
\hline
NCO$^-$+Li & 5.83 & 15666 & 5.44 & 13486 &  82.1 &  197 &  766 &  4252 &  476 & 104 \\
NCO$^-$+Na & 6.55 & 10702 & 6.18 &  8945 &  83.2 &  200 & 1020 &  4307 &  532 & 116 \\
NCO$^-$+K  & 7.17 & 10422 & 6.78 &  8965 & 145.4 &  349 & 2361 &  7526 &  804 & 174 \\
NCO$^-$+Rb & 7.41 & 10003 & 7.02 &  8622 & 159.8 &  384 & 3404 &  8269 &  901 & 194 \\
NCO$^-$+Cs & 7.68 & 10138 & 7.28 &  8780 & 197.7 &  475 & 5314 & 10236 & 1087 & 233 \\
NCO$^-$+Mg & 6.25 & 10733 & 5.90 &  8727 &  35.9 & 86.2 &  436 &  1858 &  418 & 88.0 \\
NCO$^-$+Ca & 6.67 & 14986 & 6.29 & 12966 &  78.5 &  188 & 1530 &  4061 &  713 & 152 \\
NCO$^-$+Sr & 6.94 & 14970 & 6.55 & 13004 &  99.6 &  239 & 2347 &  5156 &  854 & 181 \\
NCO$^-$+Ba & 7.16 & 16352 & 6.79 & 14231 & 138.4 &  332 & 4392 &  7164 & 1073 & 228 \\
\hline
C$_2$H$^-$+Li & 5.05 & 17041 & 6.69 &  2445 &  82.1 &  -417 &  971 &  982 &  621 & 68.3 \\
C$_2$H$^-$+Na & 5.74 & 12158 & 7.38 &  1930 &  83.2 &  -422 & 1228 &  995 &  687 & 75.3 \\
C$_2$H$^-$+K  & 6.42 & 11524 & 7.73 &  2740 & 145.4 &  -738 & 2724 & 1738 & 1039 & 114 \\
C$_2$H$^-$+Rb & 6.66 & 11047 & 7.95 &  2795 & 159.8 &  -810 & 3803 & 1910 & 1157 & 126 \\
C$_2$H$^-$+Cs & 6.94 & 11143 & 8.19 &  3089 & 197.7 & -1003 & 5808 & 2364 & 1392 & 152 \\
C$_2$H$^-$+Mg & 5.46 & 11870 & 8.36 &   854 &  35.9 &  -182 &  526 &  429 &  516 & 56.0 \\
C$_2$H$^-$+Ca & 5.92 & 15923 & 7.90 &  1892 &  78.5 &  -398 & 1725 &  938 &  896 & 97.5 \\
C$_2$H$^-$+Sr & 6.19 & 15916 & 7.93 &  2357 &  99.6 &  -505 & 2596 & 1191 & 1074 & 117 \\
C$_2$H$^-$+Ba & 6.41 & 17513 & 7.92 &  3314 & 138.4 &  -702 & 4738 & 1654 & 1353 & 147 \\
\hline
C$_4$H$^-$+Li & 7.54 & 14885 & 11.24 &  473 &  82.1 &  -801 & 1684 &  7030 &  989 & 320 \\
C$_4$H$^-$+Na & 8.24 & 10332 & 11.46 &  465 &  83.2 &  -811 & 1949 &  7119 & 1096 & 351 \\
C$_4$H$^-$+K  & 8.93 &  9825 & 11.09 &  791 & 145.4 & -1418 & 3985 & 12442 & 1656 & 529 \\
C$_4$H$^-$+Rb & 9.19 &  9367 & 11.18 &  873 & 159.8 & -1558 & 5188 & 13670 & 1847 & 586 \\
C$_4$H$^-$+Cs & 9.47 &  9441 & 11.27 & 1066 & 197.7 & -1929 & 7523 & 16922 & 2222 & 702 \\
C$_4$H$^-$+Mg & 7.98 &  9585 & 11.89 &  296 &  35.9 &  -350 &  837 &  3072 &  832 & 254 \\
C$_4$H$^-$+Ca & 8.43 & 13536 & 12.16 &  481 &  78.5 &  -765 & 2406 &  6713 & 1439 & 447 \\
C$_4$H$^-$+Sr & 8.70 & 13564 & 12.16 &  577 &  99.6 &  -971 & 3460 &  8523 & 1723 & 535 \\
C$_4$H$^-$+Ba & 8.92 & 15049 & 11.96 &  769 & 138.4 & -1350 & 5938 & 11843 & 2169 & 676 \\
\end{tabular}
\end{ruledtabular}
\end{table*}

Figure~\ref{fig:geom} shows the equilibrium geometries of molecular anions calculated with the aug-cc-pCV5Z basis set.
They agree {very well} with the experimental internuclear equilibrium distances of 1.822$\,$bohr and 2.224$\,$bohr for OH$^-$~\cite{RosenbaumJCP86} and CN$^-$~\cite{BradforthJCP93}, respectively, while the N-C and C-O experimental equilibrium distances in NCO$^-$ are 2.211$\,$bohr and 2.381$\,$bohr~\cite{BradforthJCP93}. 
The C-C and C-H experimental equilibrium distances in C$_2$H$^-$ are 2.40$\,$bohr and 2.02$\,$bohr~\cite{AtojiJCP72}, whereas no experimental data exists for C$_4$H$^-$.

Table~\ref{tab:anions} presents the permanent electric dipole moments, permanent electric quadrupole moments, parallel and perpendicular components of the static electric dipole polarizability, and vertical electron detachment energies of investigated molecular anions calculated at the equilibrium geometries with two basis sets: aug-cc-pVQZ and aug-cc-pCV5Z. The former basis set is used in calculations of intermolecular interactions, whereas the latter one is the largest basis set, which can be used to obtain molecular properties of considered anions.
Values obtained with these two basis sets agree with each other within 1-3\%. This confirms that already the smaller basis set can provide an accurate description of the considered systems.  
The experimental molecular electron affinities determined by photoelectron spectroscopy are 14741.01(3)$\,$cm$^{-1}$,  31150(30)$\,$cm$^{-1}$,  29110(30)$\,$cm$^{-1}$, 23950(50)$\,$cm$^{-1}$, and 28700(120)$\,$cm$^{-1}$ for OH, CN, NCO, C$_2$H, and C$_4$H, respectively~\cite{RienstraCR02}. They agree with calculated vertical electron detachment energies of corresponding molecular anions within 63-1300$\,$cm$^{-1}$ that correspond to an error of 0.4-4.4$\,\%$. The calculated permanent electric dipole moments also agrees with previous theoretical results within a few percent~\cite{WernerJCP83,LeeJCP85}.    
The above agreement suggests that the employed method can correctly describe the considered anions.

\begin{figure*}[t!]
\begin{center}
\includegraphics[width=\textwidth]{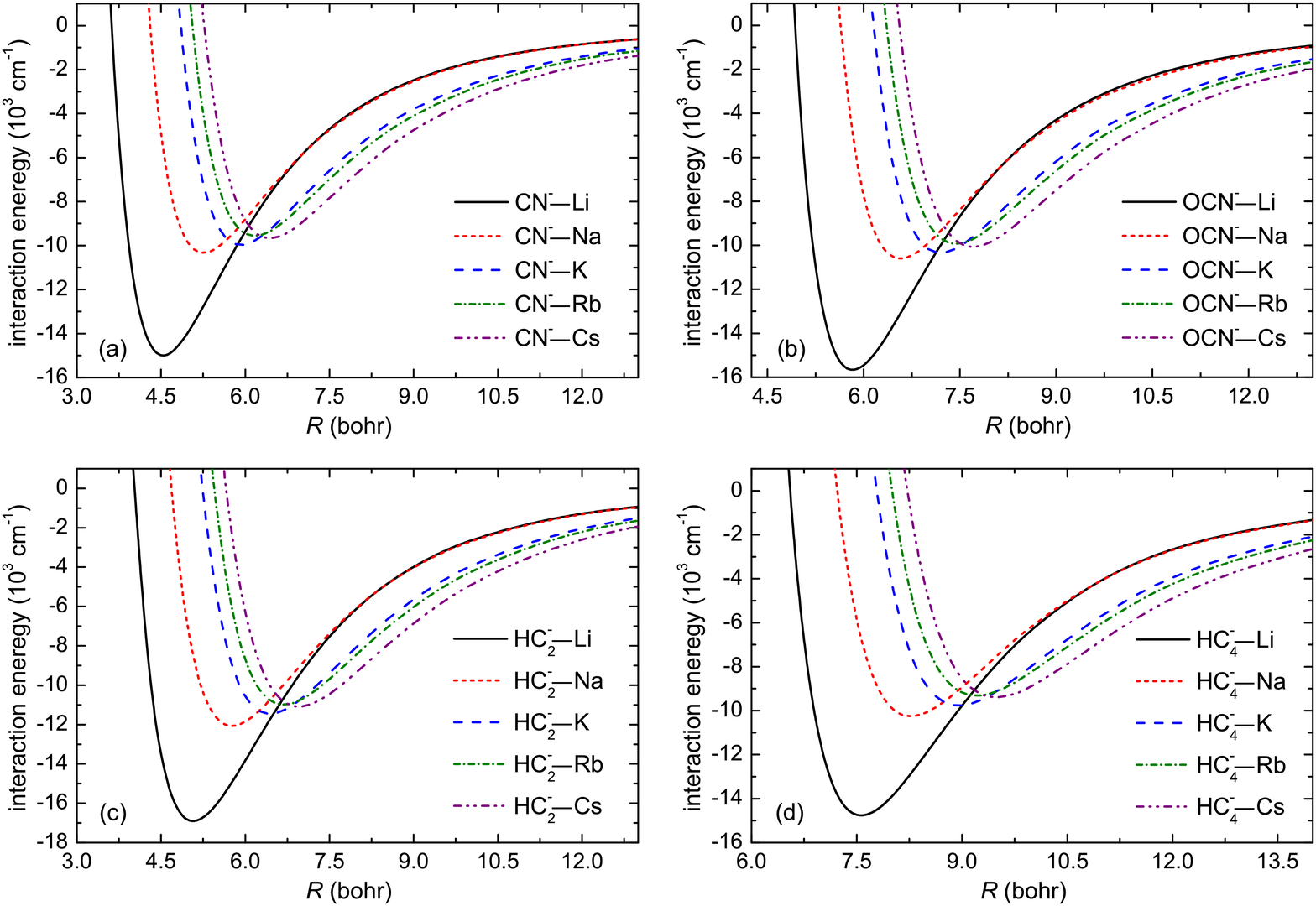}
\end{center}
\caption{One-dimensional cuts through the ground-state potential energy surfaces of molecular anions interacting with alkali-metal atoms at the linear arrangement.}
\label{fig:cuts_AM}
\end{figure*}

\begin{figure*}[t!]
\begin{center}
\includegraphics[width=\textwidth]{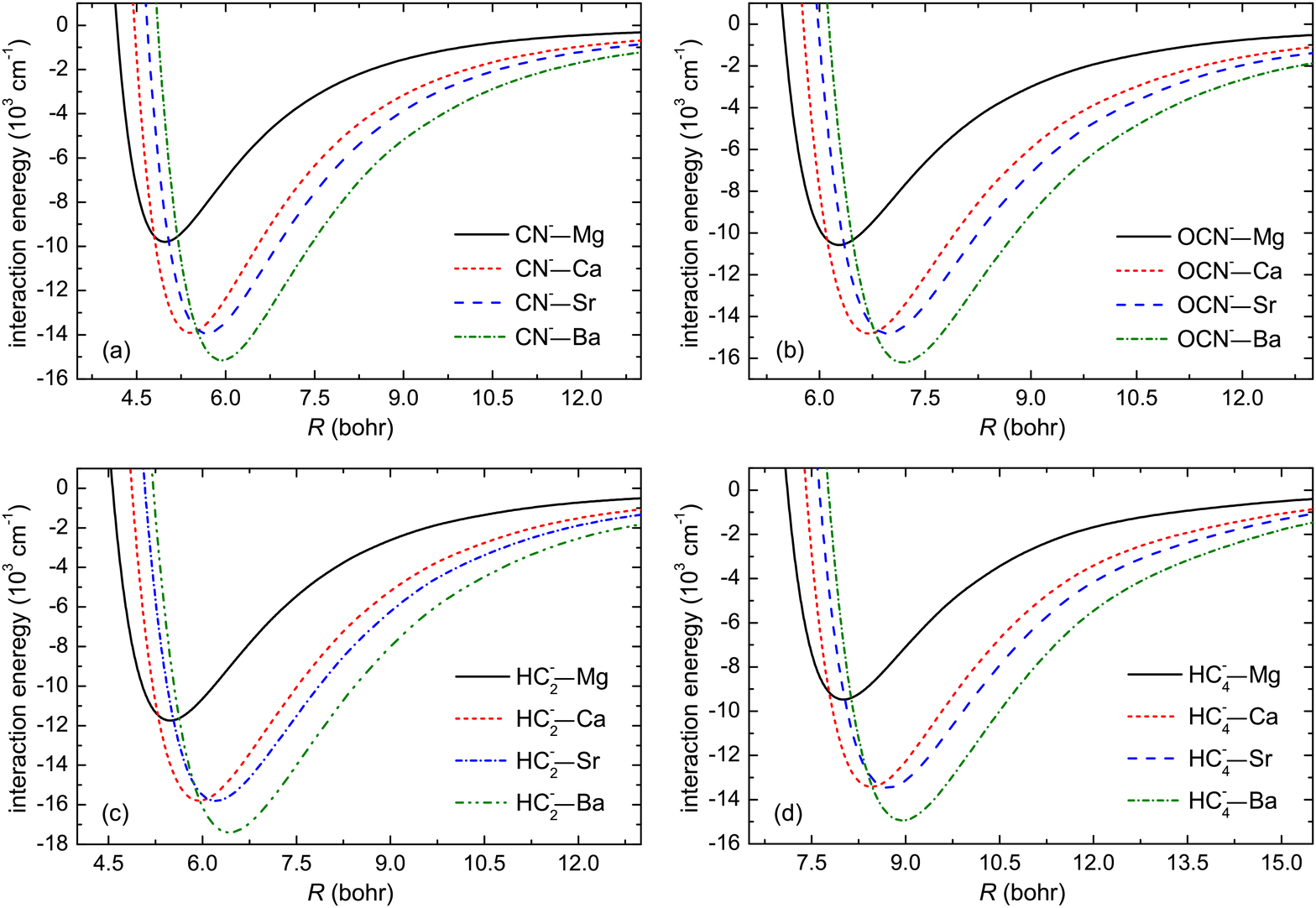}
\end{center}
\caption{One-dimensional cuts through the ground-state potential energy surfaces of molecular anions interacting with alkaline-earth-metal atoms at the linear arrangement.}
\label{fig:cuts_AEM}
\end{figure*}

Table~\ref{tab:atoms} collects the static electric dipole and quadrupole polarizabilities, ionization potentials, electron affinities, and the lowest $S-P$ excitation energies of alkali-metal and alkaline-earth-metal atoms. Present theoretical values are compared with the most accurate available experimental or theoretical data.
The calculated static electric dipole and quadrupole polarizabilities coincide with previous data within 0.1-5.7$\,$a.u.~and 9-314$\,$a.u.~that correspond to an error of 0.1-2.2$\,\%$ and 0.6-3.7$\,\%$, respectively.
The ionization potentials and the lowest $S-P$ excitation energies agree with experiential data within 23-255$\,$cm$^{-1}$ and 7-190$\,$cm$^{-1}$ that is 0.05-0.6$\,\%$ and 0.05-1$\,\%$, respectively. The electron affinities of alkali-metal atoms coincide with experiential data within 13-85$\,$cm$^{-1}$ (0.3-2$\,\%$). The Mg$^-$ anion is unstable and other alkaline-earth-metal anions are weakly bound~\cite{AndersenJPCRD99}, therefore calculations of electron affinities for alkaline-earth-metal atoms are less accurate. Nevertheless, overal agreement between calculated atomic properties and the most accurate available experimental or theoretical data is very good. 

The overall high accuracy of the calculated properties of molecular anions and atoms confirms that the employed CCSD(T) method, basis sets, and energy-consistent pseudopotentials properly treat relativistic effects and reproduce correlation energy, while being close to converged in the size of the basis function set. Thus, the used methodology should also provide an accurate description of intermolecular interactions and energetics of chemical reactions investigated in the next subsections.  
Based on the above and our previous experience, we estimate the total uncertainty of the calculated potential
energy surfaces at the global minimum to be of the order of 200-500$\,$cm$^{-1}$ that corresponds to 2-5\% of the interaction energy. The uncertainty of the long-range interaction coefficients is of the same order of magnitude.

\subsection{Potential energy surfaces}

Figures~\ref{fig:cuts_AM} and~\ref{fig:cuts_AEM} present one-dimensional cuts through the ground-state potential energy surfaces at the linear arrangement of the CN$^-$, NCO$^-$, C$_2$H$^-$, and C$_4$H$^-$ molecular anions interacting with the Li, Na, K, Rb, Cs alkali-metal and Mg, Ca, Sr, Ba alkaline-earth-metal atoms, respectively. For the NCO$^-$, C$_2$H$^-$, and C$_4$H$^-$ molecular anions the presented minima are global.    
The equilibrium intermolecular distances $R_e$ and well depths $D_e$ corresponding to the presented arrangement and for the second linear geometry are collected in Table~\ref{tab:int}. The leading long-range induction and dispersion interaction coefficients are also reported in Table~\ref{tab:int}.

An inspection of the potential energy curves for different anions presented in Figs.~\ref{fig:cuts_AM} and~\ref{fig:cuts_AEM} reveals remarkable similarities. 
All potential energy curves show a smooth behavior with well-defined minima. The pattern of shapes and relative positions of curves with different atoms at the linear geometry is very similar for all investigated anions.  The similarity at large internuclear distances is not surprising since the leading long-range induction interactions are determined by the polarizability of atoms and the charge of anions. The similarity at small internuclear distances shows that all investigated anions behave similar regarding the short-range electrostatic and exchange interactions with alkali-metal and alkaline-earth-metal atoms.
This observation suggests that, on one hand, the potential energy surfaces obtained for some anion-atom system can be used to describe other anion-atom systems by proper scaling energy and length, and on the other hand, the effective potential energy surfaces can be generated by combining the long-range multipole expansion of the intermolecular interaction energy with some short-range repulsion term intended to reproduce typical biding energies.

\begin{figure*}[t!]
\begin{center}
\includegraphics[width=\columnwidth]{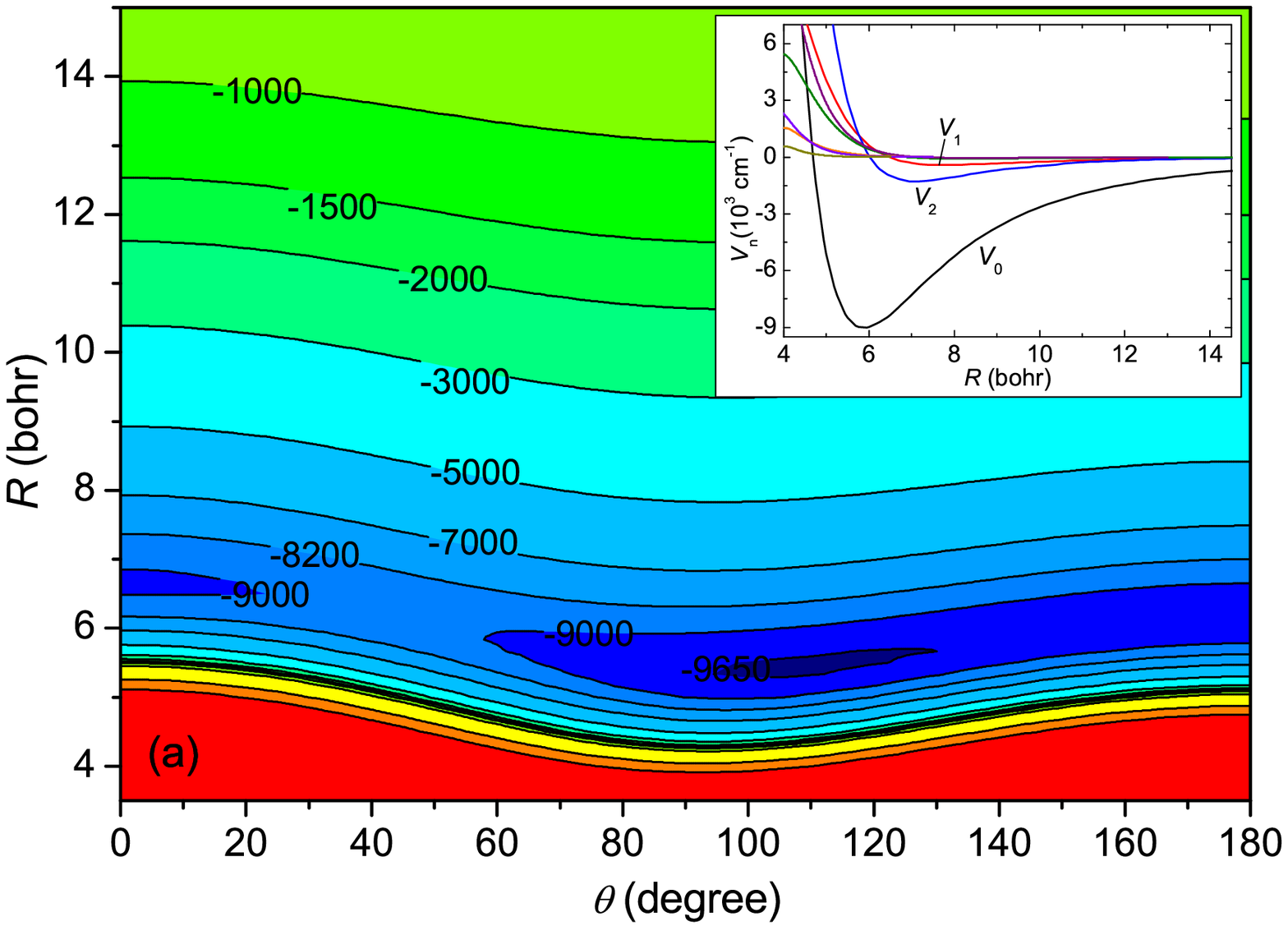}
\includegraphics[width=\columnwidth]{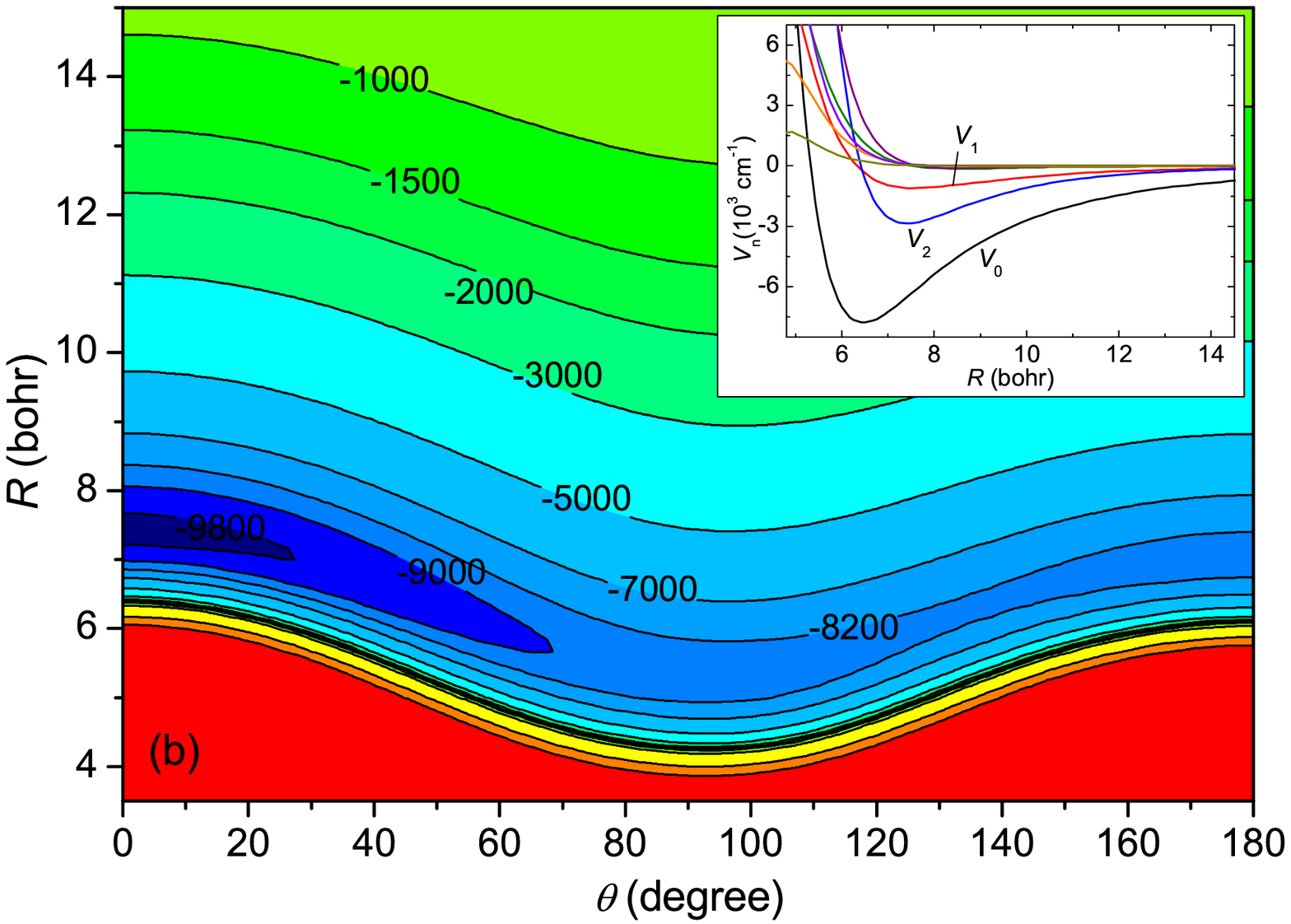}
\includegraphics[width=\columnwidth]{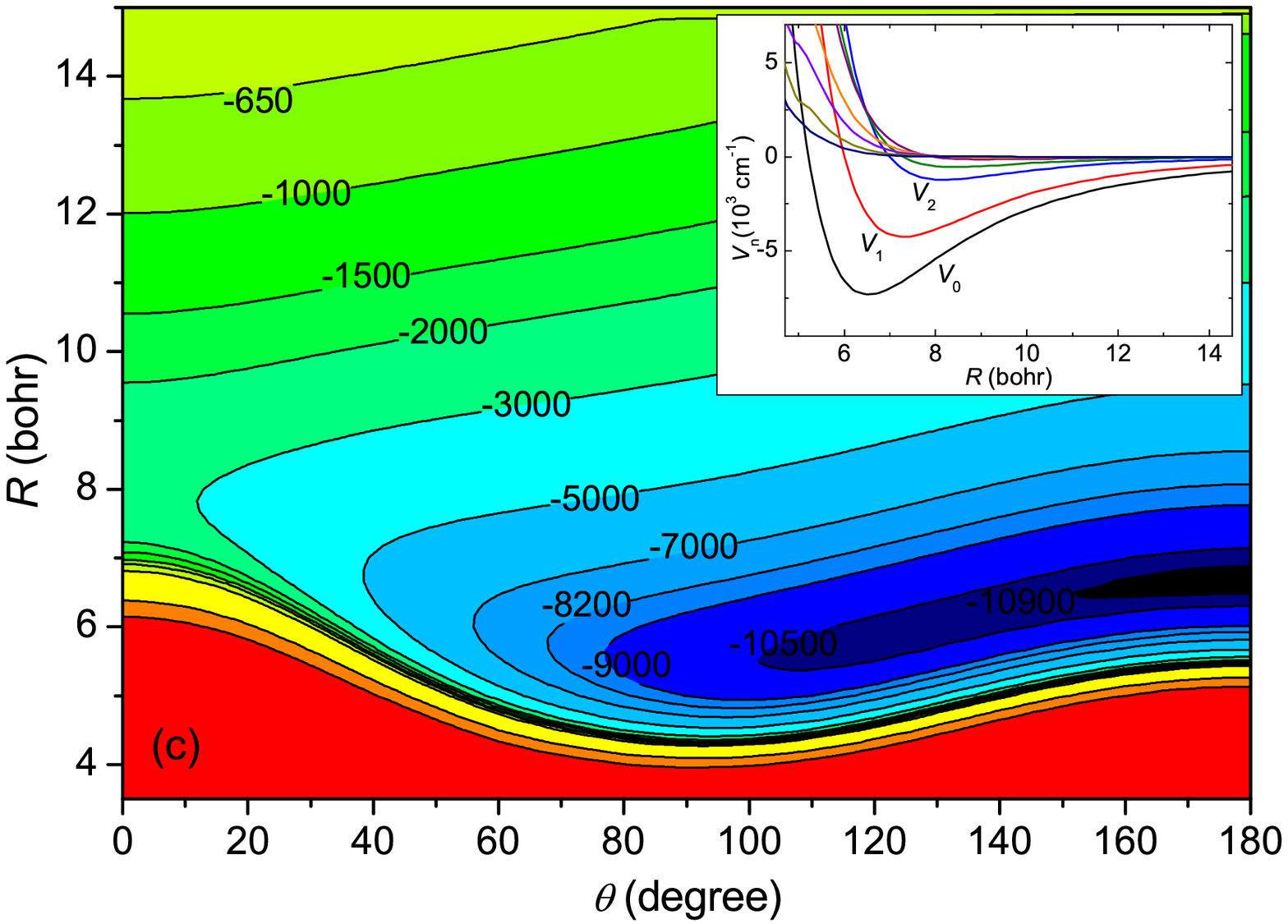}
\includegraphics[width=\columnwidth]{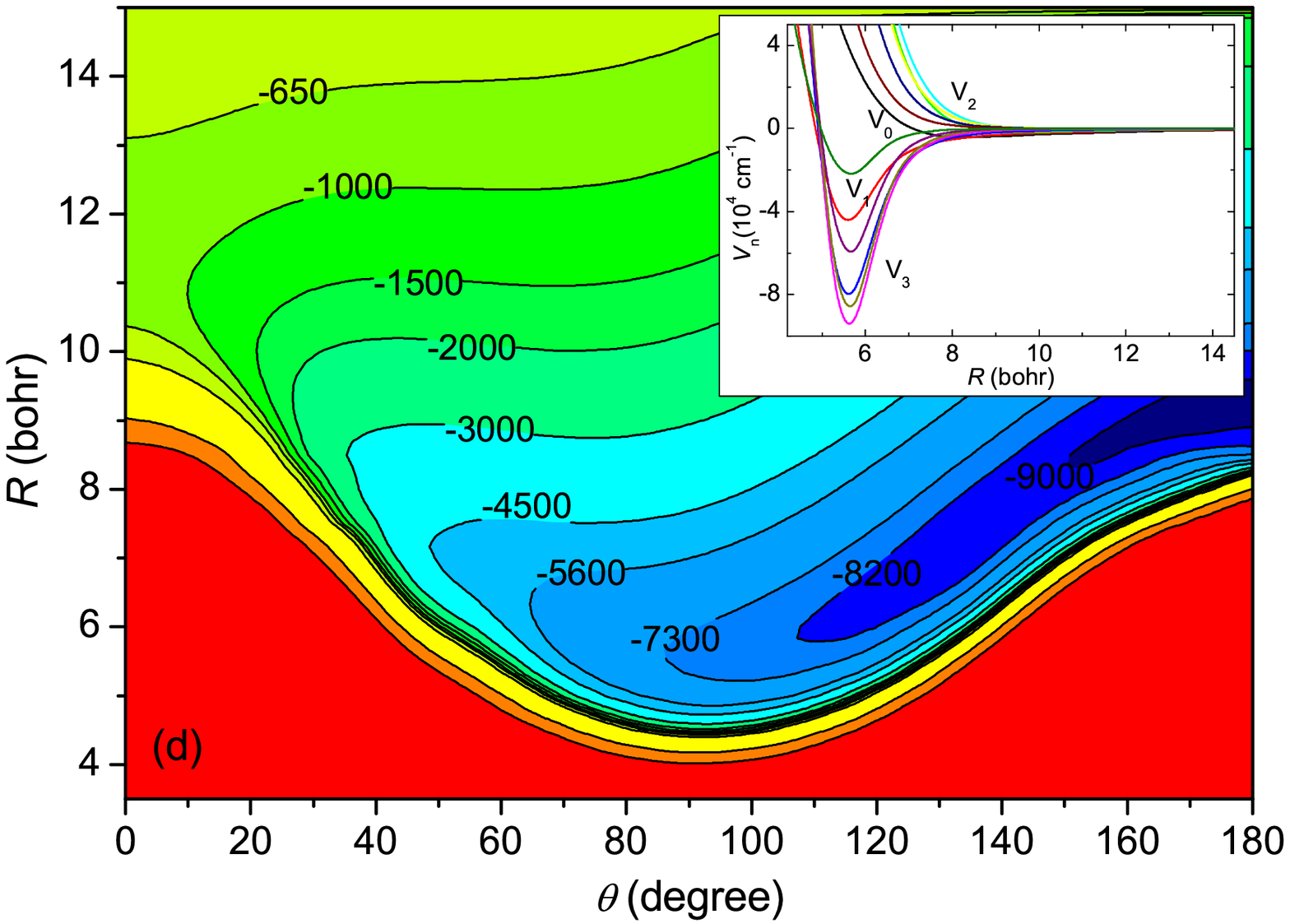}
\end{center}
\caption{The ground-state potential energy surfaces for: (a) CN$^-$+Rb, (b) NCO$^-$+Rb, (c) C$_2$H$^-$+Rb, and (d) C$_4$H$^-$+Rb. Insets show the corresponding Legendre components. }
\label{fig:surfs}
\end{figure*}
 
For all investigated anion-atom systems, the equilibrium intermolecular distance increases with the mass of an atom (e.g.~from 3.18$\,$bohr for OH$^-$+Li to 4.40$\,$bohr for OH$^-$+Ba, and from 7.54$\,$bohr C$_4$H$^-$+Li to 8.92$\,$bohr for C$_4$H$^-$+Ba). Instead, the well depth decreases with the mass of an atom for alkali-metal atoms and increases with the mass of an atom for alkaline-earth-metal atoms. The different trends for alkali-metal atoms as compared with alkaline-earth-metal atoms are typical for non-covalent interactions and were also observed for interactions of these atoms with chromium~\cite{TomzaPRA13a,DengCPL16}, europium~\cite{TomzaPRA14}, and closed-shell~\cite{GueroutPRA10,BruePRA13} atoms. It can be explained by the fact that the formal order of the chemical bond is equal to half for the complexes of closed-shell species with alkali-metal atoms and zero for
the ones with alkaline-earth-metal atoms. For this reason the former ones are chemically bound whereas the latter
ones are stabilized by the induction and dispersion interactions only.

Among anion--alkali-metal-atom systems, the largest binding energy is for complexes with the lithium atom (the well depth is between 14,885$\,$cm$^2$ for C$_4$H$^-$+Li and 23,917$\,$cm$^2$ for OH$^-$+Li), while the binding energies for complexes with other alkali-metal atoms are 30$\,$\% smaller.
Among anion--alkaline-earth-metal-atom systems, the largest binding energy is for complexes with the barium atom (the well depth is between 15,049$\,$cm$^2$ for C$_4$H$^-$+Ba and 27,552$\,$cm$^2$ for OH$^-$+Ba) and the smallest binding energy is for complexes with the magnesium atom (the well depth is between 9,585$\,$cm$^2$ for C$_4$H$^-$+Mg and 20,283$\,$cm$^2$ for OH$^-$+Mg).

Figure~\ref{fig:surfs} presents the ground-state potential energy surfaces for the CN$^-$, NCO$^-$, C$_2$H$^-$, and C$_4$H$^-$ molecular anions interacting with the Rb atom. For the CN$^-$ anion the global minimum is at the non-linear geometry ($R_e$=5.44$\,$bohr, $\theta_e$=$108^{\circ}$, $D_e=$9723$\,$cm$^{-1}$), whereas for other anions the global minima are at the linear geometry in the configuration where the Rb atom approaches the molecular anion from its more charged side. The calculated potential energy surfaces have two minima for CN$^-$ and one minimum for other anions and are strongly anisotropic with the anisotropy increasing with the size of the molecular anion. For the CN$^-$ and OCN$^-$ anions the second anisotropic Legendre term is larger than the first one, $V_2(R)>V_1(R)$, which means that the potential energy surfaces are relatively symmetric with respect to exchange the C and N or O and N atoms in the CN$^-$ or OCN$^-$ anions, respectively.  
For the C$_2$H$^-$ anion the first anisotropic Legendre term $V_1(R)$ is almost as large as the isotropic one $V_0(R)$, whereas for the C$_4$H$^-$ anion the PES is heavily anisotropic with several anisotropic terms larger than the isotropic one. The large dipole moment, related to the localization of the charge on the ending carbon atom, and increasing size of the considered molecular anions are responsible for the observed large anisotropy. The topology of potential energy surfaces for the considered anions interacting with other alkali-metal and alkaline-earth-metal atoms is similar. 

The potential energy surfaces were previously investigated for OH$^-$+Rb~\cite{GonzalezEPJD08} and CN$^-$+Rb/Sr~\cite{MidyaPRA16} anion-atom systems. The authors of the former reference used approach similar to the present one and obtained very similar results, whereas the authors of the latter reference used larger basis sets but their results also agree within 0.5$\,\%$ with the present ones.

{ Unfortunately, at the moment, even the most accurate PESs do not allow one to predict accurately the scattering lengths for collisions between
many-electron atoms and molecules. Nevertheless, the general characteristics of cold elastic and inelastic collisions, and thus prospects for sympathetic cooling, can be learned from scattering calculations by tuning the scattering lengths around values typical for ion-atom interactions and testing them against the uncertainty of PESs~\cite{ZuchowskiPRA09,TomzaPRA15a}.
Furthermore, in the future the presented PESs can be corrected using the scattering data from experiments, thereby allowing for fully quantitative predictions.
}

\subsection{Prospects for chemical reactions}

The prospects for sympathetic cooling and applications of molecular ions immersed into ultracold atomic gases can be jeopardized by possible chemical reactions, on one hand, however cold and controlled chemical reactions in these systems can be an interesting subject of study on its own, on the other hand.

In cold mixtures of molecular anions and atoms several types of possible chemical reactions induced by intermolecular interactions or external fields during collisions can be envisioned.

(i) The spontaneous radiative charge transfer
\begin{equation}\label{eq:RCT}
A^- + M \to A + M^- + h\nu\,,
\end{equation}
where the electron is spontaneously transfered from a molecular anion $A^-$ to an atom $M$ emitting a photon of an energy $h\nu$.  
This process is possible if the electron affinity (EA) of a neutral atom is larger than the electron detachment energy (ED) of a molecular anion. The electron detachment energy of a molecular anion is usually very close to the electron affinity of a corresponding neutral molecule. The energy of a produced photon is equal to the difference of the EA and ED energies. 

{ The spontaneous non-radiative charge transfer can also be possible for the same energetic conditions if electronic states associated with $A^- + M$ and $A + M^-$ thresholds form an avoided crossing or conical intersection at short internuclear distances.}

(ii) The spontaneous radiative association
\begin{equation}\label{eq:RA}
A^- + M \to (MA)^- + h\nu\,,
\end{equation}
where a molecular anion $A^-$ and an atom $M$ spontaneously form an ionic complex $(MA)^-$ emitting a photon with an energy $h\nu$. Such a process { driven by the transition between two electronic states} is possible when the reaction \eqref{eq:RCT} is energetically allowed or when the interaction energy in a complex $(MA)^-$ is greater or equal to the missing difference of the EA and ED energies. { The spontaneous radiative association is also possible (but very unlikely) for all polar complexes $(MA)^-$ driven by the transition between rovibrational states of one electronic state.}

(iii) The photo-induced charge transfer
\begin{equation}
A^- + M + h\nu \to A^- + M^* (A^{*-} + M) \to A + M^-\,,
\end{equation}
where the spontaneous radiative charge transfer is energetically not allowed and the missing energy is introduced by exiting an molecular anion $A^-$ or an atom $M$ with a laser field $h\nu$. The photon of a smaller energy $h\nu'<h\nu$ can be produced in such a process, too. Once the charge transfer is photo-induced, the spontaneous radiative association (ii) is also possible. 

(iv) The electron detachment association (associative electron detachment)
\begin{equation}
A^- + M \to MA + e^-\,,
\end{equation}
where a molecular anion $A^-$ and an atom $M$ form a neutral complex $MA$ and at the same time the electron is detached from the system. Such a process is possible when the interaction energy in a neutral complex $MA$ is greater or equal to the electron detachment energy of a molecular anion.

(v) The collision-induced isomerization
\begin{equation}
ABC^- + M \to ACB^- + M\,,
\end{equation}
where one metastable isomer of a molecular anion $ABC^-$ is transformed into another isomer $ACB^-$ as a result of a collision with an atom $M$. In such a scenario the anion-atom interactions provide the energy needed to overcome the isomerization energy barrier, thus an atom serves as a catalyzer. 

(vi) The proper chemical reaction with the rearrangement of atoms between reactants
\begin{equation}\label{eq:true}
AB^- + M \to A^- + MB \,\, (A + MB^-)\,,
\end{equation}
where an atom $B$ is transfered from a molecular anion $AB^-$ to an atom $M$ forming a neutral molecule $MB$ or a molecular anion $MB^-$. Such a process is possible
when the dissociation energy of $MB$ or $MB^-$ is larger than the dissociation energy of $AB^-$. 

Processes (iv), (v), and (vi), if energetically forbidden, can also be photo-induced by exiting an molecular anion  or an atom with a laser field.

For all investigated anion-atom systems the spontaneous radiative { and non-radiative} charge transfer, reaction (i), is energetically not allowed because the electron detachment energy of the OH$^-$, CN$^-$, NCO$^-$, C$_2$H$^-$, and C$_4$H$^-$ molecular anions (cf.~Tab.~\ref{tab:anions}) is much larger than the electron affinity of the Li, Na, K, Rb, Cs, Mg, Ca, Sr, and Ba atoms (cf.~Tab.~\ref{tab:atoms}). The charge transfer reaction can be easiest photo-induced, reaction (iii), for collisions between the OH$^-$ molecular anion and alkali-metal atoms. For these systems the lowest $S-P$ excitation of alkali-metal atoms provides a sufficient amount of energy. For other anion-atom systems the higher excitation of atoms or anions is needed.

The above observed stability of molecular anions against spontaneous radiative charge transfer in collisions with alkali-metal and alkaline-earth-metal atoms is typical and should also be expected for other anions. In contrast, most of molecular cations are expected to experience radiative-charge-transfer losses in collisions with alkali-metal and alkaline-earth-metal atoms because of the relatively low ionization energy of these atoms~\cite{TomzaPRA15b}.

The spontaneous radiative association, reaction (ii), is energetically allowed only for collisions between the OH$^-$ molecular anion and alkali-metal atoms. This process was already investigated for OH$^-$+Rb both experimentally~\cite{DeiglmayrPRA12} and theoretically~\cite{ByrdPRA13}. For other investigated anion-atom systems the interaction energy (cf.~Tab.~\ref{tab:int}) is not large enough to overcome the electron detachment energy of molecular anions (cf.~Tab.~\ref{tab:anions}). However, the lowest $S-P$ excitation of alkali-metal and alkaline-earth-metal atoms provides a sufficient amount of energy to photo-induce reaction (ii) with the C$_2$H$^-$ anion and for several other anion-atom systems. 

The electron detachment association, reaction (iv), may be energetically allowed only for collisions between the OH$^-$ molecular anion and alkali-metal atoms, because the electron detachment energy of other molecular anions is too large. This reaction can potentially be also photo-induced, however more detailed studies are needed for specific anion-atom systems.

The collision-induced isomerization, reaction (v), is feasible only for the NCO$^-$ cyanate anion which can exist in a metastable isomeric form as the CNO$^-$ fulminate anion~\cite{HolsboerJCSD70}. The activation energy (reaction barrier) for the $\mathrm{CNO}^-\to\mathrm{NCO}^-$ isomerization is predicted to be around 16,000$\,$cm$^{-1}$~\cite{YoungshangJCP97,LeonardJCP10} that is of the same order of magnitude as the interaction energy between the NCO$^-$ anion and alkali-metal and alkaline-earth-metal atoms (cf.~Tab.~\ref{tab:int}). Thus, the anion-atom interaction energy can potentially be sufficient to overcome the isomerization energy barrier, however more detailed studies are needed.

The proper chemical reactions with breaking and formation of bonds are not expected for collisions between the OH$^-$, CN$^-$, and NCO$^-$ molecular anions and considered atoms, because these anions are strongly bound as compared to species consisting of alkali-metal or alkaline-earth-metal atoms with C, N, and H atoms~\cite{}. Example complete analysis of chemical reaction channels for the Rb+OH$^-$ system is presented in Ref.~\cite{ByrdPRA13}. For the C$_2$H$^-$ and C$_4$H$^-$ molecular anions, breaking the C-H bound, $\mathrm{C}_n\mathrm{H}^-\to \mathrm{C}_n+\mathrm{H}^-$ or $\mathrm{C}_n\mathrm{H}^-\to \mathrm{C}_n^-+\mathrm{H}$, should be the easiest. Unfortunately, the dissociation energies for those reactions are almost 60,000$\,$cm$^{-1}$ and 40,000$\,$cm$^{-1}$ (64,000$\,$cm$^{-1}$ and 70,000$\,$cm$^{-1}$) for C$_2$H$^-$ (C$_4$H$^-$), respectively, whereas the dissociation energies of neutral or anionic hydrides of alkali-metal and alkaline-earth-metal atoms do not exceed 30,000$\,$cm$^{-1}$~\cite{GeumJCP01,AymarJPB12}. Thus, the proper chemical reactions are also energetically not allowed for collisions between the C$_2$H$^-$ and C$_4$H$^-$ molecular anions and alkali-metal and alkaline-earth-metal atoms.

\section{SUMMARY AND CONCLUSIONS}
\label{sec:summary}

After many spectacular successes in the field of ultracold atoms, the scientific community has drawn its attention to the research on ultracold molecules. Recently, ultracold gases of diatomic molecules have been produced and explored. The next emerging goal is the preparation of polyatomic molecules at ultralow temperatures and the first experiments have been launched. 
Molecular ions are easier to prepare, trap, and detect as compared to neutral molecules. They are also important in many areas of chemistry ranging from organic and inorganic chemistry to astrochemistry. 
Therefore polyatomic molecular ions are promising systems to start investigating cold polyatomic dynamics and chemical reactions at the quantum level.  

Here, we have investigated the electronic structure and intermolecular interactions of several molecular anions (OH$^-$, CN$^-$, NCO$^-$, C$_2$H$^-$, C$_4$H$^-$) with alkali-metal (Li, Na, K, Rb, Cs) and alkaline-earth-metal (Mg, Ca, Sr, Ba) atoms. We have calculated and characterized the potential energy surfaces, long-range induction and dispersion interaction coefficients, and possible channels of chemical reactions and their control by using state-of-the-art \textit{ab initio} techniques: the coupled cluster method restricted to single, double, and noniterative triple excitations, CCSD(T), combined with the large Gaussian basis sets and small-core energy-consistent pseudopotentials.

We have shown that most of the considered anion-atom systems are stable against chemical reactions and charge transfer processes which however can be induced by exciting atoms or anions with the laser field. Thus the present work opens the ways for collisional studies of linear polyatomic ions immersed in ultracold atomic gases and their applications in controlled chemistry, precision measurements, and quantum simulations.

{The first experiments combining diatomic molecular ions with ultracold atoms have used a Paul trap to trap ions~\cite{HallPRL12,RellergertNature13,DeiglmayrPRA12}. This trapping technique is indispensably associated with the micromotion of ions induced by the rf field. In such a scenario, sympathetic cooling can be prevented and ion-atom collisions can lead to heating, e.g.~if atoms are heavier than ions~\cite{CetinaPRL12}. One can potentially avoid this kind of heating by using an optical dipole trap to trap ions, as was demonstrated for atomic ions~\cite{SchneiderNatPhot10} and suggested for diatomic anions~\cite{YzombardPRL15}. 
The possible detection schemes for molecular anions include the laser-induced fluorescence or molecular-ion trap-depletion spectroscopy~\cite{ChenPRA11}. However, more detailed studies of both trapping and detection techniques are needed for the considered here anions and molecular ions in general.
}

The present study of the electronic structure is the first step towards the evaluation of prospects for sympathetic cooling and controlled chemistry of linear polyatomic anions with ultracold alkali-metal or alkaline-earth-metal atoms.
This work also establishes the computational scheme for the future \textit{ab initio} investigations of intermolecular interactions in other polyatomic anion-atom systems relevant for ultracold physics or chemistry and can serve as the benchmark for investigations of more challenging polyatomic cation-atom systems.
In the future, the obtained potential energy surfaces and long-range interaction coefficients will be employed in time-independent scattering calculations for both elastic and inelastic collisions at low and ultralow temperatures and their control with magnetic and laser fields.

\begin{acknowledgments}
Financial support from the National Science Centre Poland (2015/19/D/ST4/02173) and the PL-Grid Infrastructure is gratefully acknowledged. 
\end{acknowledgments}

\bibliography{anions}

\end{document}